\documentclass[prb,showpacs]{revtex4}
\usepackage{epsfig}
\usepackage{graphicx}
\usepackage{amsmath}
\usepackage{bm}
\usepackage{amssymb}
\usepackage{color}

\newif\ifold             \oldtrue            
\def\ba{\begin{eqnarray}}
\def\ea{\end{eqnarray}}

\newcommand{\be}{\begin{equation}}
\newcommand{\ee}{\end{equation}}

\begin{document}


\title{Broken-symmetry states and phase diagram of the lowest Landau level in bilayer graphene}
\date{\today}

\author{E. V. Gorbar}
\affiliation{Bogolyubov Institute for Theoretical Physics, 03680, Kiev, Ukraine}

\author{V. P. Gusynin}
\affiliation{Bogolyubov Institute for Theoretical Physics, 03680, Kiev, Ukraine}

\author{Junji Jia}
\affiliation{Department of Applied Mathematics, University of Western Ontario, London,
Ontario N6A 5B7, Canada}

\author{V. A. Miransky}
\affiliation{Department of Applied Mathematics, University of Western Ontario, London,
Ontario N6A 5B7, Canada}

\begin{abstract}

Broken-symmetry quantum Hall (QH) states with filling factors $\nu=0, \pm 1, \pm 2, \pm 3$
in the lowest Landau level in bilayer graphene are analyzed by solving the gap equation
in the random phase approximation. It is shown that in the plane of electric and magnetic
fields,  the critical line, which separates the spin and layer polarized phases at $\nu=0$,
extends to the $\nu=\pm 1$ QH states. The amplitudes of the gaps in the $\nu= \pm 1,
\pm 3$, and $\nu= \pm 2$ QH states are significantly smaller than the amplitude
of the $\nu=0$ gap, due to the separate filling of the $n=0$ and $n=1$ orbital Landau
levels and the negative contribution of the Hartree term, respectively. It is shown that
those values of the external electric field where the conductance is not quantized
correspond to the minima of the gaps.
\end{abstract}


\maketitle
\section{Introduction}
\label{1}

Bilayer graphene is a new material with unique
properties.\cite{McC,Nov,Hen,CN} The possibility of inducing a tunable bandgap
by top-bottom gates voltage makes it very promising for applications in various
electronic devices.

Recent experiments in bilayer graphene \cite{FMY,Weitz,Martin,Freitag,Zh,SKim}
revealed the generation of energy gaps in a magnetic field
with complete lifting of the eightfold degeneracy in the zero energy (lowest) Landau
level (LLL), which leads to new quantum Hall states with filling factors
$\nu=0,\pm1,\pm2,\pm3$. While in Refs.~\onlinecite{FMY,Weitz,Martin,Freitag} suspended
bilayer graphene was used, bilayer graphene samples deposited on Si${\mbox O_2}$/Si
substrates were utilized in Refs.~\onlinecite{Zh,SKim}.
Because suspended bilayer graphene is much cleaner than that on a substrate, the new
quantum Hall states in the former start to develop at essentially smaller magnetic fields
than in the latter. The theory of the quantum Hall (QH) effect in bilayer graphene has
been studied in
Refs.~\onlinecite{Bar,Abergel,Sh,NCD,Nandkishore1,GGM1,GGM2,Nandkishore2,Falko,Kharitonov}.

It was revealed in Ref.~\onlinecite{FMY} that the energy gaps scale linearly with magnetic
field $B$ in bilayer graphene. It is unlike the case of monolayer graphene where a
$\sqrt{B}$ scaling for the gaps takes place.\cite{Zhang2006,Jiang2007} As was
suggested in Refs.~\onlinecite{Nandkishore1,GGM1,GGM2}, a strong screening produced by the Coulomb
interaction is responsible for this modification of the scaling in bilayer.\cite{footnote} The
physics underlying this effect is the following.\cite{GGM1,GGM2} Due to a nonrelativistic
like dispersion relations for quasiparticles,\cite{McC} the polarization
function in bilayer graphene in a magnetic field is strongly enhanced as compared to
the case of monolayer graphene. In particular,
it is proportional to the large mass of the quasiparticles $m \simeq 10^{-2}m_e$.
Such a strong screening radically changes the form of the interaction and
leads to the linear scaling.

Another interesting phenomenon in the $\nu = 0$ QH state in bilayer graphene
is the phase transition between the spin polarized (ferromagnetic) phase and the layer
polarized one in the $BE_{\perp}$-plane, where $E_{\perp}$ is an electric field orthogonal
to the bilayer planes. It was analyzed in theoretical studies in
Refs.~\onlinecite{GGM1,GGM2,Nandkishore2,Falko} and observed in experiments in
Refs.~\onlinecite{Weitz,SKim,Freitag}.

The main conclusion of the analysis in Refs.~\onlinecite{GGM1,GGM2} was that these two phases are separated
by the critical line $\tilde{\Delta}_0=\mu_B\,B+e^2d/l^2$, where $\mu_B$ is the Bohr magneton,
$l=\sqrt{\hbar c/eB}$ is the magnetic length, $\tilde{\Delta}_0 = eE_{\perp}d/2$ is the
top-bottom gates voltage imbalance, and $d \simeq 0.35$nm is the distance between the graphene
layers. This critical line is qualitatively consistent with the experimental results in
Refs.~\onlinecite{Weitz,SKim,Freitag}.

Only the $\nu=0$ QH state was analyzed in Refs.~\onlinecite{GGM1,GGM2,Nandkishore2,Falko}.
As to the QH states with $\nu = \pm 1, \pm 2, \pm 3$, the main experimental results obtained in
Refs.~\onlinecite{Weitz,Martin} are the following: a) For $\nu= \pm 1$, there
are two phases separated by approximately the same critical line as that in the $\nu=0$ QH state.
b) There is only one phase for $\nu= \pm 2$ and $\nu = \pm 3$ QH states.
c) The $\nu=0$ gap is approximately $30-40\%$ larger than the $\nu= \pm 2$
one and significantly (by factor $10$) exceeds the $\nu=\pm 1$ gap. d) For $\nu= 0$, $\nu=\pm 1$,
$\nu=\pm 2$, and $\nu=\pm 3$, the conductance is quantized except at particular values of the
electric field $E_\perp$.

In this paper we extend the analysis of Refs.~\onlinecite{GGM1,GGM2} beyond the neutral point
and describe the QH states with the filling factors $\nu= \pm 1, \pm 2, \pm 3$. As will be
shown below, the results of the analysis reproduce correctly the main characteristics of
the experimental data.

The paper is organized as follows. In Sec. \ref{2}, the Hamiltonian of the model,
its symmetries, and order parameters are described. In Sec. \ref{3}, by using the
Baym-Kadanoff formalism, the gap equation for the quasiparticle propagator
including the polarization function is derived. In Sec. \ref{4}, the properties of the
solutions of the gap equation and the phase diagram of the LLL are described.
In Sec. \ref{5}, we compare our results with experiment. In Sec. \ref{6}, the main
results of the paper are summarized. In Appendix we comment on the bare splitting
of the Landau levels with orbital indices 0 and 1 induced by a bias electric field.

\section{Model}
\label{2}

{We will utilize the same model for describing the low-energy electronic excitations
as in Refs.~\onlinecite{GGM1,GGM2}. It can be derived from the more accurate four-band
model, which takes into account the sombrero  shape of the band structure for zero magnetic field.
\cite{McC,CN,Chakraborty}}
The free part of the effective low energy Hamiltonian of bilayer graphene is:\cite{McC}
\be
H_0 = - \frac{1}{2m}\int
d^2x\Psi_{Vs}^+(x)\left( \begin{array}{cc} 0 & (\pi^{\dagger})^2\\ \pi^2 & 0
\end{array} \right)\Psi_{Vs}(x), \label{free-Hamiltonian}
\ee
where $\pi=\hat{p}_{x_{1}}+i\hat{p}_{x_{2}}$ and the canonical momentum
$\hat{\mathbf{p}} = -i\hbar\bm{\nabla}+ {e\mathbf{A}}/c$ includes the
vector potential $\mathbf{A}$ corresponding to the external magnetic field
$\mathbf{B}$. Without magnetic field, this Hamiltonian generates the spectrum
$E=\pm \frac{p^2}{2m}$, where $m= \gamma_1/2v_{F}^2$ with the Fermi velocity $v_F
\simeq c/300$ and $\gamma_1 \approx 0.34-0.40$ eV. The two component spinor
field $\Psi_{Vs}$ carries the valley $(V= K, K^{\prime})$ and spin
$(s = +, -)$ indices. We will use the standard convention:
$\Psi_{Ks}^T=(\psi_A{_1}, \psi_B{_2})_{Ks}$
whereas $\Psi_{K^{\prime}s}^T = (\psi_B{_2},
\psi_A{_1})_{K^{\prime}s}$. Here $A_1$ and $B_2$ correspond to
those sublattices in the layers 1 (top) and 2 (bottom), respectively, which, according to
Bernal $(A_2-B_1)$ stacking,
are relevant for the low energy dynamics.

The Zeeman and Coulomb interactions plus a top-bottom gates voltage imbalance
$\tilde{\Delta}_0$ in bilayer graphene are described as
(henceforth we will omit indices $V$ and $s$ in the field $\Psi_{Vs}$):
{\ba
H_{\rm int}&=& Z\hspace{-1.0mm} {\rm \int}\hspace{-1.0mm} d^2x \Psi^+(x)
\sigma^3\Psi(x)\nonumber +
\frac{1}{2}\int \hspace{-1.0mm}d^2xd^2x^{\prime}\left\{
V(x-x^{\prime})\left[\rho_1(x)\rho_1(x^{\prime})+
\rho_2(x)\rho_2(x^{\prime})\right]\hspace{-1.0mm} +
2V_{12}(x-x^{\prime})\rho_1(x)\rho_2(x^{\prime})\right\}\\
&+&
\tilde{\Delta}_0 \int d^2x \Psi^+(x)\xi\tau_3\Psi(x)-\frac{\tilde{\Delta}_0}{m\gamma_{1}}
\int d^2x \Psi^+(x)\xi\left( \begin{array}{cc}\pi^{\dagger}\pi&0\\
0&-\pi\pi^{\dagger}\end{array}\right)\Psi(x)\,,
\label{interaction1}
\ea
}
where $\sigma^3$ is a spin matrix, $Z \equiv \mu_{B}B=0.67\,B[{\mbox T}]{\mbox K}$ is the Zeeman
energy, the Pauli matrix $\tau^3$ in the
voltage imbalance term acts on layer components,
and $\xi = \pm 1$ for the valleys $K$ and $K^{\prime}$, respectively. The potential $V(x)$
describes the intralayer interactions and, therefore,
coincides with the bare potential in monolayer graphene whose Fourier transform
is given by $\tilde{V}(k)={2\pi e^2}/{\kappa k}$, where $\kappa$ is the dielectric constant.
The potential $V_{12}$ describes
the interlayer electron interactions. Its Fourier transform is
$\tilde{V}_{12}(k)=({2\pi e^2}/{\kappa})({e^{-kd}}/{k})$.
The two-dimensional charge densities $\rho_1(x)$ and $\rho_2(x)$ are:
\be
\rho_1(x)=\Psi^+(x)P_1\Psi(x)\,,\quad \rho_2(x)=\Psi^+(x)P_2\Psi(x)\,,
\label{density}
\ee
where $P_1=(1+\xi\tau^3)/2$ and $P_2=(1-\xi\tau^3)/2$ are projectors
on states in the layers 1 and 2, respectively. When the polarization
effects are taken into account, the potentials $V(x)$ and $V_{12}(x)$ are replaced
by effective interactions $V_{\mbox{\scriptsize eff}}(x)$ and $V_{\mbox{\scriptsize 12eff}}(x)$. {The last term
in Eq.(\ref{interaction1}) leads to the splitting of the Landau levels with orbital
indices $n=0$ and $n=1$.\cite{McC,CN,Chakraborty}
As is discussed in Appendix  below, for
the values of the magnetic field $B < 10$T, which are relevant for the experiment
with suspended bilayer graphene\cite{Weitz,Martin} and the present analysis, the contribution
of this term is small. On the other hand, it could be relevant for stronger magnetic fields.
Because of that, this term is omitted in the analysis of the gap equations
in the main text, although we discuss some interesting features this term could lead to in
the case of stronger magnetic fields in Secs. \ref{4a} - \ref{4d} and Appendix.

If both the Zeeman and $\tilde{\Delta}_0$ terms are ignored, the Hamiltonian
$H = H_0 + H_{\rm int}$, with $H_0$ and $H_{\rm int}$ in Eqs. (\ref{free-Hamiltonian}) and
(\ref{interaction1}),
possesses the symmetry $G = U^{(K)}(2)_S \times U^{(K^{\prime})}(2)_S\times
Z_{2V}^{(+)}\times Z_{2V}^{(-)}$, where $U^{(V)}(2)_S$ defines the $U(2)$ spin
transformations in a fixed valley $V = K, K^{\prime}$, and $Z_{2V}^{(s)}$
describes the valley transformation $\xi \to -\xi$ for a fixed spin $s = \pm$.\cite{GGM1,GGM2}
The Zeeman interaction lowers this symmetry down to $G_2 \equiv U^{(K)}(1)_{+}
\times U^{(K)}(1)_{-} \times U^{(K^{\prime})}(1)_{+} \times U^{(K^{\prime})
}(1)_{-} \times Z_{2V}^{(+)}\times Z_{2V}^{(-)}$, where $U^{(V)}(1)_{s}$ is the
$U(1)$ transformation for fixed values of both valley and spin.
Including the $\tilde{\Delta}_0$ term lowers the $G_2$ symmetry
further down to the $\bar{G}_2 \equiv
U^{(K)}(1)_{+}\times U^{(K)}(1)_{-} \times U^{(K^{\prime})}(1)_{+} \times
U^{(K^{\prime})}(1)_{-}$.

The dynamics in the integer QH effect in bilayer graphene is intimately
connected with dynamical breakdown of the $G$ and $G_2$ symmetries.
Two sets of the order parameters describing their breakdown were considered in
Refs.~\onlinecite{GGM1,GGM2}. The first set consists of the quantum Hall ferromagnetism (QHF)
order parameters:\cite{QHF}
{\begin{eqnarray}
\label{mu}
\mu_s:\quad
\langle{\Psi^\dagger_s\Psi_s}\rangle &=&
\langle{\psi_{K  A_1 s}^\dagger\psi_{K A_1 s}
+\psi_{K^{\prime} A_1 s}^\dagger\psi_{K^{\prime}A_1 s}
 +\psi_{K B_2 s}^\dagger\psi_{K B_2 s}
+ \psi_{K^{\prime}B_2 s}^\dagger\psi_{K^{\prime} B_2 s}}\rangle\,,\\
\label{mu-tilde}
\tilde{\mu}_{s}:\quad
\langle{\Psi^\dagger_s\xi\Psi_s}\rangle &=&
\langle{\psi_{K  A_1 s}^\dagger\psi_{K A_1 s}
 + \psi_{K B_2 s}^\dagger\psi_{K
B_2 s} - \psi_{K^{\prime}B_2 s}^\dagger\psi_{K^{\prime} B_2 s}
- \psi_{K^{\prime} A_1 s}^\dagger\psi_{K^{\prime}A_1 s}}\rangle\,.
\end{eqnarray}
The order parameter (\ref{mu}) is the charge density for a fixed spin
whereas the order parameter (\ref{mu-tilde}) determines the charge-density
imbalance between the two valleys. The corresponding chemical potentials
are $\mu_s$ and $\tilde{\mu}_s$, respectively.

The second set consists of the magnetic catalysis (MC) order parameters,\cite{MC}
i.e., the Dirac $\tilde{\Delta}_s$ and Haldane $\Delta_s$ mass terms:
\begin{eqnarray}
\label{delta}
\Delta_s: \quad
\langle{\Psi^\dagger_s\tau_3\Psi_s}\rangle &=&
\langle{\psi_{K  A_1 s}^\dagger\psi_{K A_1 s}
 -\psi_{K B_2 s}^\dagger\psi_{K
B_2 s}+ \psi_{K^{\prime}B_2 s}^\dagger\psi_{K^{\prime} B_2 s} -\psi_{K^{\prime} A_1 s}^\dagger
\psi_{K^{\prime}A_1 s}}\rangle\,,\\
\label{delta-tilde}
\tilde{\Delta}_{s}:\quad
\langle{\Psi^\dagger_s\xi\tau_3\Psi_s}\rangle &=&
\langle{\psi_{K  A_1 s}^\dagger\psi_{K A_1 s}
+\psi_{K^{\prime} A_1 s}^\dagger\psi_{K^{\prime}A_1 s}
 -\psi_{K B_2 s}^\dagger\psi_{K
B_2 s} - \psi_{K^{\prime}B_2 s}^\dagger\psi_{K^{\prime} B_2 s}}\rangle\,.
\end{eqnarray}
Clearly, the order parameter (\ref{delta}) describes a charge density wave in both the $K$
and $K^{\prime}$ valleys. While this order parameter
preserves the $G_2$ symmetry, it is odd under time reversal.\cite{Hald} On the other hand,
the order parameter (\ref{delta-tilde}),
connected with the conventional Dirac mass $\tilde{\Delta}_s$,
determines the charge-density
imbalance between the two layers.\cite{McC}} Like the QHF order parameter
(\ref{mu-tilde}), this mass term completely breaks the $Z_{2V}^{(s)}$
symmetry and is even under time reversal. It is important that
in both monolayer and bilayer graphene, these two sets of the order parameters necessarily
coexist\cite{GGM1,GGM2,Gorb1} and are produced even at the weakest repulsive interactions between
electrons (magnetic catalysis\cite{catal,Khvesh,Gorb}).
The essence of this phenomenon
is an effective reduction by two units of the spatial dimension in the
electron-hole pairing in the LLL with energy $E=0$\,.

Let us also emphasize that unlike a
spontaneous breakdown of continuous symmetries, a spontaneous breakdown of
the discrete valley symmetry $Z_{2V}^{(s)}$, with the order parameters
$\langle{\Psi^\dagger_s\xi\Psi_s}\rangle$
and $\langle{\Psi^\dagger_s\xi\tau_3\Psi_s}\rangle$,
is not forbidden by the Mermin-Wagner theorem at finite
temperatures in a planar system.\cite{MW} Also, because the valley and layer indices
are equivalent in the LLL,\cite{McC} these order parameters also describe a breakdown
of the symmetry between the top and bottom layers.

Note that because of the
Zeeman interaction, the $SU^{(V)}(2)_S$ is explicitly broken, leading to a
spin gap. This gap could be dynamically strongly enhanced \cite{Aban}. In that
case, a quasispontaneous breakdown of the $SU^{(V)}(2)_S$ takes place.
The corresponding ferromagnetic phase is described by the chemical potential
$\mu_3 = (\mu_+ - \mu_-)/2$, corresponding to the
QHF order parameter $\langle\Psi^\dagger\sigma_3\Psi\rangle$, and by the mass
$\Delta_3 = (\Delta_+ - \Delta_-)/2$ corresponding to
the MC order parameter
$\langle\Psi^\dagger\tau_3\sigma_3\Psi\rangle$.

Recall also that in bilayer graphene, the LLL includes both the $n=0$ and $n=1$ LLs,
if the Coulomb interaction is ignored.\cite{McC} Therefore, in the LLL approximation,
there is an approximate  orbital symmetry $Z_{2L}$ connected with a $n=0 \to n=1$ transformation.

\section{Gap equation}
\label{3}

The gap equation for the full propagator $G$ in the LLL approximation in bilayer graphene
was derived in Ref.~\onlinecite{GGM2}.
In the derivation, the Baym-Kadanoff (BK) formalism\cite{BK} was utilized. In this section, we
describe the main features of the gap equation. Note that because a characteristic scale in the bilayer
dynamics in a magnetic field is the cyclotron energy $\hbar \omega_c \simeq 2.19 B[{\mbox T}] \mbox{meV}$,
the applicability of the LLL approximation implies that the LLL energy gaps should be smaller
than $\hbar \omega_c$. As we will see, this condition is fulfilled in bilayer graphene.

We will analyze the gap equation for the full quasiparticle propagator
$G$ with the order parameters introduced in Eqs. (\ref{mu})-(\ref{delta-tilde}). The mean field approximation
with the polarization function calculated in the random phase approximation (RPA) will be used. The
corresponding two-loop BK effective action is a functional of the full quasiparticle propagator $G$ and
it has the form:\cite{GGM2}
\begin{equation}
\Gamma(G)=\Gamma_0(G)+\Gamma_2(G)\,,\quad\quad\quad
\Gamma_0(G)={-i}\,\mbox{Tr}\left[\mbox{Ln} G^{-1} +S^{-1}G-1\right]\,,
\label{CJT-action-all}
\end{equation}
where $S$ is the free Green's function, $\Gamma_0$ is the free part, and $\Gamma_2$ takes into account the
interaction effects in the first order of perturbation theory,
 \begin{eqnarray}
\Gamma_2(G)&=&- \int d^{3}ud^3u^{\prime}\left\{\frac{1}{2}\mbox{tr}
\left[G(u,u^{\prime})G(u^{\prime},u)\right]V_{\mbox{\scriptsize eff}}(u-u^{\prime}) +
\mbox{tr}\,[P_1\,G(u,u^{\prime})\,P_2\,G(u^{\prime},u)\,]\,V_{\mbox{\scriptsize IL}}(u-u^{\prime})\right.\nonumber\\
&&
\left.-\frac{1}{2}\mbox{tr}\left[G(u,u)\right]\mbox{tr}\left[G(u^{\prime},
u^{\prime})\right]V_{\mbox{\scriptsize eff}}(u-u^{\prime})-\mbox{tr}\,[\,P_1\,G(u,u)]\,
\mbox{tr}\,[\,P_2\,G(u^{\prime},u^{\prime})\,]\,V_{\mbox{\scriptsize IL}}(u-u^{\prime})
\right\}\,.
\label{Gamma2}
\end{eqnarray}
Here $u \equiv (t,\mathbf{r})$, $t$ is the time coordinate, $\mathbf{r} = (x,y)$, and
$V_{\mbox{\scriptsize IL}}(u)=V_{\mbox{\scriptsize 12eff}}(u)-V_{\mbox{\scriptsize eff}}(u)$
is a combination of the interlayer and intralayer interactions (recall that
the polarization contributions are included in the potentials $V_{\mbox{\scriptsize eff}}(u)$
and $V_{\mbox{\scriptsize 12eff}}(u)$). Note that while here the trace
\rm{Tr}, the logarithm, and the product $S^{-1}G$ are taken in the functional sense, the trace
\rm{tr} runs over layer, valley and spin indices. The stationary condition $\delta \Gamma /\delta G = 0$
leads to the gap equation for the quasiparticle propagator $G$ in mean field approximation.

We will use the Landau gauge for a two dimensional vector potential,
$\mathbf{A}=(0,Bx)$, where $B$ is the component of the
magnetic field $\mathbf{B}$ orthogonal to the $xy$ plane of graphene. Then,
the free Green's function $S(u_1,u_2)$ can be written as a product of a translation
invariant part $\tilde{S}(u_1 -u_2)$ times the Schwinger phase factor \cite{catal}
\be
S(u_1,u_2)=\exp\left(-i\frac{(x_1 + x_2)(y_1 -y_2)}{2l^{2}}\right)\tilde{S}(u_1 - u_2).
\ee
Similarly, we can separate the Schwinger phase factor from the translation invariant part in the full propagator
\be
G(u_1,u_2)=\exp\left(-i\frac{(x_1 + x_2)(y_1 -y_2)}{2l^{2}}\right)\tilde{G}(u_1 - u_2)\,.
\label{coordinate-space-propagator}
\ee
For the LLL with the orbital numbers $n = 0, 1$,
the translation invariant part of the time Fourier transform of the free propagator takes a simple form:
\be
\tilde{S}_{\xi s}(\mathbf{r};\omega)=\frac{1}{2\pi l^{2}}\exp\left(-\frac{\mathbf{r}^{2}}
{4l^{2}}\right)\left[L_{0}\left(\frac{\mathbf{r}^{2}}{2l^{2}}\right)+
L_{1}\left(\frac{\mathbf{r}^{2}}{2l^{2}}\right)\right]{S}_{\xi s}(\omega)P_{-}
\label{free-propagator-LLL}
\ee
with
\be
{S}_{\xi s}(\omega)=\frac{1}{\omega+\mu_0-sZ+\xi\tilde{\Delta}_{0} + i\delta{\rm sgn\omega}}\,,
\ee
where $\mu_0$ is the electron chemical potential, $Z$ is the Zeeman energy, and the projector
$P_{-}=(1-\tau_{3})/2$.

{Motivated by expression (\ref{free-propagator-LLL}) for $\tilde{S}_{\xi s}(\mathbf{r};\omega)$,
we will use the following ansatz for the full propagator with the
parameters $\mu_s(n)$, $\tilde{\mu}_{s}(n)$, $\Delta_s(n)$,
and $\tilde{\Delta}_s(n)$ related to the order parameters in Eqs. (\ref{mu}) -- (\ref{delta-tilde}):
\be
\tilde{G}_{\xi s}(\mathbf{r};\omega)=\frac{1}{2\pi l^{2}}\exp\left(-\frac{\mathbf{r}^{2}}
{4l^{2}}\right)\left[G_{\xi 0 s}(\omega)L_{0}\left(\frac{\mathbf{r}^{2}}{2l^{2}}\right)+
G_{\xi 1 s}(\omega)L_{1}\left(\frac{\mathbf{r}^{2}}{2l^{2}}\right)\right]P_{-},
\label{ansatz}
\ee
where
\be
G_{\xi n s}(\omega)=\frac{1}{\omega-E_{\xi ns}+ i\delta{\rm sgn\omega}}
\label{G-expression}
\ee
and
\be E_{\xi ns}= -(\mu_{s}(n)+ \Delta_{s}(n))+ \xi(\tilde{\mu}_{s}(n) -
\tilde{\Delta}_{s}(n)),\quad n=0,1,
\label{disp}
\ee
are the energies of the LLL states depending on the parameters $\mu_{s}(n),
\tilde{\mu}_{s}(n),\Delta_{s}(n), and \tilde{\Delta}_{s}(n)$. Note that because
for the LLL states only the component $\psi_{B_2s}$ $(\psi_{A_1s})$ of the wave
function at the $K (K^{\prime})$ valley is nonzero, their energies depend only
on the eight independent combinations of the QHF and MC parameters shown in
Eq. (\ref{disp}).\cite{footnote1}
Our goal is to find the energies $E_{\xi ns}$ from the gap equation.

It is convenient to solve the gap equation in momentum space.
Then, the Fourier transform  $\tilde{V}_{\mbox{\scriptsize eff}}(\omega,k)$
of $V_{\mbox{\scriptsize eff}}(u)$ is:\cite{GGM2}}
\be
\tilde{V}_{\mbox{\scriptsize eff}}(\omega,k)=\frac{2\pi e^2}{\kappa}\,\frac{1}{ k
+\frac{4\pi e^2}{\kappa}\Pi (\omega,{\bf k}^2)}
\label{V-D}
\ee
with $\Pi(\omega,k^{2})\equiv \Pi_{11}(\omega,\mathbf{k})+\Pi_{12}(\omega,\mathbf{k})$,
where the polarization function $\Pi_{ij}$ describes electron densities
correlations on the layers $i$ and $j$ in a magnetic field. For the polarization function
$\Pi(\omega,k^{2})$ we use the expression in the RPA approximation modified by the presence
of a quasiparticle bare gap $\tilde{\Delta}_{0}$ term (see Eq.(A24) in Ref. \onlinecite{GGM2}).
As to the potential $V_{\mbox{\scriptsize IL}}(u)$,
the gap equation contains its Fourier transform only  at zero frequency and momentum:\cite{GGM2}
\be
\tilde{V}_{\mbox{\scriptsize IL}}(\omega=0,{k}=0)=-\frac{2\pi e^{2}d}{\kappa}\,.
\label{interlayer-interaction1}
\ee
The inclusion of the polarization effects is crucial for ensuring the linear scaling
in bilayer graphene. We utilize the frequency independent order parameters $\mu,\tilde{\mu},
\Delta,\tilde{\Delta}$ and the static approximation for the polarization function will be used,
$\Pi(\omega,{\bf k}^2) \to \Pi(0,{\bf k}^2)$.\cite{footnote2}
The explicit expression for the polarization function
calculated in the RPA approximation can be found in Ref.~\onlinecite{GGM2}.

Using the ansatz (\ref{ansatz}), one finds that the gap equation $\delta \Gamma /\delta G = 0$
leads to the following system of equations  for the energies $E_{\xi sn}$
(for details, see Ref.~[\onlinecite{GGM2}]):
\ba
-E_{\xi 0s}&=&{\mu}_{0}-s Z+\xi\tilde{\Delta}_{0}-\frac{\hbar^{2}}{2ml^{2}}\left[{\rm sgn}
\left(E_{\xi 0s}\right)I_{1}(x)+{\rm sgn}\left(E_{\xi 1s}\right)I_{2}(x)\right]\nonumber\\
&+&\frac{1}{4\pi l^{2}}\left[(A_{1}+A_{2})\,\tilde{V}_{\mbox{\scriptsize eff}}(0)+
\left(\frac{1-\xi}{2}A_{2}+\frac{1+\xi}{2}A_{1}\,
\right)\,\tilde{V}_{\mbox{\scriptsize IL}}(0)\right],
\label{system-eq1}\\
-E_{\xi 1s}&=&{\mu}_{0}- s Z+\xi\tilde{\Delta}_{0}-\frac{\hbar^{2}}{2ml^{2}}\left[{\rm
sgn}
\left(E_{\xi 0s}\right)I_{2}(x)+{\rm sgn}\left(E_{\xi 1s}\right)I_{3}(x)\right]\nonumber\\
&+&\frac{1}{4\pi l^{2}}\left[(A_{1}+A_{2})\,\tilde{V}_{\mbox{\scriptsize eff}}(0)+
\left(\frac{1-\xi}{2}A_{2}+\frac{1+\xi}{2}A_{1}
\right)\,\tilde{V}_{\mbox{\scriptsize IL}}(0)\right],
\label{system-eq2}
\ea
where the integrals $I_{i}(x)$ are
\begin{equation}
I_i(x)=\int_0^{\infty} \frac{dy\,f_i(y)\,e^{-y}}{\kappa \sqrt{xy} + 4\pi
\tilde{\Pi}(y)}
\label{integralsI-i}
\end{equation}
with $f_i(y)=(1,\,y,\,(1-y)^2)$ for $i=1,2,3$, respectively.
Here the dimensionless
variable $x=2\hbar^{4}/e^{4}m^{2}l^{2}=(4\hbar\omega_{c}/\alpha^{2}\gamma_{1})
(v_{F}/c)^{2}\simeq 0.003 B[{\mbox T}]$, where $\alpha=1/137$ is the
fine-structure constant and we used the values $\gamma_{1}=0.39
\mbox{eV}$, $\hbar\omega_{c}=\hbar^{2}/ml^{2}=2.19 B[{\mbox T}]\mbox{meV}$,
$v_{F}=8.0\times10^{5}$m/s (see Ref.~\onlinecite{McC}).
The quantities $A_1$ and $A_2$ are  $A_1 =
\sum_{n,s}\,\mbox{sgn}(\,E_{-1 ns})$\,, $A_2 =\sum_{n,s}\,\mbox{sgn}(\,E_{1 ns})$.

The terms in the first and second square brackets on the right hand side
of Eqs. (\ref{system-eq1}) and (\ref{system-eq2}) describe the exchange and Hartree
contributions, respectively. In the latter, the terms with the factor
$(A_1 + A_2)/{4\pi l^2}$ are proportional to ${\rm tr}[G(0)]$,\cite{GGM2} i.e.,
the density of charge carriers. There are other Hartree contributions, such as
those taking into account the background charge of ions in graphene, the charge
in the substrate, etc., which are not included in the equations. Due to the overall
neutrality of the system, all these contributions should exactly cancel. As a result,
only the Hartree terms with the factor $(A_1 - A_2)/{4\pi l^2}$ survive. They
describe the capacitor like interactions in bilayer graphene.
Thus, the final equations, taking into account the neutrality condition, are:
\begin{eqnarray}
E_{\xi 0s}&=&-{\mu}_{0}+s Z-\xi\tilde{\Delta}_{0}+\frac{\hbar^{2}}{2ml^{2}}\left[{\rm sgn}
\left(E_{\xi 0s}\right)I_{1}(x)+{\rm sgn}\left(E_{\xi 1s}\right)I_{2}(x)\right]
-\frac{\xi}{8\pi l^{2}}\,(A_{1}-A_{2})\,\tilde{V}_{\mbox{\scriptsize IL}}(0)\,,
\label{new_system-eq1}\\
E_{\xi 1s}&=&-{\mu}_{0}+ s Z-\xi\tilde{\Delta}_{0}+\frac{\hbar^{2}}{2ml^{2}}\left[{\rm sgn}
\left(E_{\xi 0s}\right)I_{2}(x)+{\rm sgn}\left(E_{\xi 1s}\right)I_{3}(x)\right]
-\frac{\xi}{8\pi l^{2}}\,(A_{1}-A_{2})\,\tilde{V}_{\mbox{\scriptsize IL}}(0)\,.
\label{new_system-eq2}
\end{eqnarray}
The filling factor
\be
\nu=-\frac{1}{2}\sum_{\xi ns} \mbox{sgn}(E_{\xi ns})
=-\frac{1}{2}(A_1+A_2)
\label{def:filling}
\ee
takes values $0, \pm 1, \pm 2, \pm 3 \pm 4$.
In general, there are many solutions of the gap equations (\ref{new_system-eq1})
and (\ref{new_system-eq2}) at a fixed filling factor $\nu$. The density
of the thermodynamical potential of the system for each
solution can be calculated by using the BK effective action (\ref{CJT-action-all})
and the fact that solutions are extrema of this action.
We find
\begin{eqnarray}
\Omega &=&-i\int\frac{d\omega}{8\pi^2l^2}\sum_{\xi ns}\left\{\frac{\omega-
\mu_0+sZ-\xi\tilde{\Delta}_0}
{\omega-E_{\xi ns} + i\delta{\rm sgn\omega}}-1\right\}\nonumber\\
&=&-\frac{1}{8\pi l^2}
\sum_{\xi=\pm}\sum_{s=\pm}\sum_{n=0,1}\left[E_{\xi ns}
-\mu_0+sZ-\xi\tilde{\Delta}_0\right]\,\mbox{sgn}(E_{\xi ns})\,.
\label{dens}
\end{eqnarray}

\section{Solutions and phase diagram in the lowest Landau level}
\label{4}


{The chemical potential is one of thermodynamic variables of the thermodynamic potential (\ref{dens}).
However, in experiments with bilayer graphene, the top and/or bottom gates control the charge
density (filling factor $\nu$) rather than the chemical potential. Therefore, it is convenient
to perform a Legendre transform and use the charge density as a thermodynamic variable.}

Usually, there is a one-to-one correspondence between the chemical potential of a system and
its charge density. However, in the case under consideration, such a correspondence is
absent. Indeed, since we consider an ideal system at zero temperature and zero LLL width,
the chemical potential can take arbitrary values within the gap between the filled and unfilled
levels. Since the LLL states corresponding to the filling factors
$\nu=0, \pm1, \pm2$, and $\pm3$ are gapped, this is a general situation in the present dynamics.

{Actually, this simplifies the analysis of these states: because as soon as the
filling factor is fixed, the chemical potential becomes an irrelevant parameter
and does not affect the energy density of the system for any solution studied.
Explicitly, using the charge density
\begin{equation}
\rho=i\int_{-\infty}^{+\infty}\frac{d\omega}{2\pi}\,\mbox{tr}\,[\,\tilde{G}(\omega;0)\,]=
-\frac{1}{4\pi l^2}
\sum_{\xi=\pm}\sum_{s=\pm}\sum_{n=0,1}\mbox{sgn}(E_{\xi ns})\,
\label{charge-density}
\end{equation}
and performing the Legendre transform, we find the following free energy density:
\begin{equation}
{\cal E}=\Omega+\mu_0\rho=-\frac{1}{8\pi l^2}
\sum_{\xi=\pm}\sum_{s=\pm}\sum_{n=0,1}\left[E_{\xi ns}
+\mu_0+sZ-\xi\tilde{\Delta}_0\right]\,\mbox{sgn}(E_{\xi ns})\,.
\label{free-energy-density}
\end{equation}
Since $-\mu_0$ additively enters $E_{\xi ns}$ defined in Eqs.(\ref{new_system-eq1})
and (\ref{new_system-eq2}), it is clear that $\mu_0$ cancels out in (\ref{free-energy-density})
for all solutions with a given filling factor.}

{Therefore, our strategy in the analysis of the gap equations is the following one:
At each fixed filling factor $\nu$ and given values of the controlling parameters $B$ and
$E_{\perp}$, we find all possible solutions of
Eqs. (\ref{new_system-eq1}) and (\ref{new_system-eq2}) with a chemical potential allowed
by the filling $\nu$, and then determine the ground
state as the solution with the lowest free energy density (\ref{free-energy-density}).}

The gap equations (\ref{new_system-eq1}) and (\ref{new_system-eq2}) form
a system of algebraic equations for the eight energies $E_{\xi 0s}$ and $E_{\xi 1s}$ of
quasiparticle levels. The solutions having the same filling factor differ in the signs of
particular energies but have the same combined values of $A_1+A_2$ (and consequently $\nu$).
Therefore in order to describe the solutions corresponding to different filling
factors $\nu$, it is sufficiently to  specify only the signs of their energies.

By pre-setting the signs of the energies, one finds that for
$\nu=0,~\pm1,~\pm2,~\pm3$, and $\pm4$, there are ${8\choose 4} =70$, ${8 \choose 3}  =56$,
${8 \choose 2}=28$, ${8 \choose 1}= 8$, and ${8 \choose 0}= 1$
solutions, respectively. Therefore there are 256 solutions all together.
In what follows, we will consider the solutions only with filling factors from 0 to +4
(the solutions with negative values of $\nu$ can be obtained from these ones
by choosing opposite values of the signs of the energies). This reduces the number of the
solutions to 163. For concreteness, we also choose positive values for a magnetic field,
$B > 0$, and nonnegative values for an applied electric field, $E_{\perp} \ge 0$.

{As is discussed below, the analysis of
Eqs. (\ref{new_system-eq1}), (\ref{new_system-eq2}) and (\ref{free-energy-density})
with a chemical potential allowed by the filling $\nu$\cite{footnote3}}
shows that for each filling factor,
there are only one or two major solutions that have the lowest energy density
and therefore are relevant for the phase diagram. Only for $\nu=1$ and $\nu=3$, and in a very
narrow region of the values of $B$ and $E_{\perp}$ (where the major phases do not exist),
there might exist some additional (marginal) phases.

In this section, the properties of the solutions of gap equations
(\ref{new_system-eq1})-(\ref{new_system-eq2}) will be described. The comparison of these
solutions with experiment will be considered in Sec. \ref{5}.

\subsection{The $\nu=0$ QH state}
\label{4a}

The solutions of the gap equations at the neutral point were analyzed in Refs.~[\onlinecite{GGM1,GGM2}].
Here we briefly describe those results.

There are two competing solutions: I) the ferromagnetic (or spin polarized (SP)) solution and II)
the ferroelectric (or layer polarized (LP))
solution. The signs of the energies $E_{\xi n s}$ in the ferromagnetic solution are
negative for the states with a spin opposite to magnetic field and positive for the states
with a spin along magnetic field:
$$
\mbox{sgn} \left( E_{{1,0,-}} \right)=-1,\,\,\mbox{sgn} \left( E_{{-1,0,-}} \right)=-1,\,\,
\mbox{sgn} \left( E_{{1,1,-}} \right) =-1,\,\,\mbox{sgn} \left( E_{{-1,1,-}} \right)=-1,
$$
\begin{equation}
\mbox{sgn} \left( E_{{1,0,+}} \right)=1,\,\,\mbox{sgn} \left( E_{{-1,0,+}} \right)=1,\,\,
\mbox{sgn} \left( E_{{1,1,+}} \right)=1,\,\,\mbox{sgn} \left( E_{{-1,1,+}} \right)=1\,.
\label{ferromagnetic}
\end{equation}
In the LP solution, the signs of the energies are correlated with the signs
of the valley index $\xi$ (recall that the valley and layer indices are equivalent in the LLL):
$$
\mbox{sgn} \left( E_{{1,0,-}} \right)=-1,\,\,\mbox{sgn} \left( E_{{1,0,+}} \right)=-1,
\mbox{sgn} \left( E_{{1,1,-}} \right)=-1,\,\,\mbox{sgn} \left( E_{{1,1,+}} \right)=-1,
$$
\begin{equation}
\mbox{sgn} \left( E_{{-1,0,-}} \right)=1,\,\,\mbox{sgn} \left( E_{{-1,0,+}} \right)=1,\,\,
\mbox{sgn} \left( E_{{-1,1,-}} \right)=1,\,\,\mbox{sgn} \left( E_{{-1,1,+}} \right) =1\,.
\label{layer-asymmetric}
\end{equation}
The analytical expressions for the energies $E_{\xi ns}$ of these two solutions can
be easily found from Eqs.(\ref{new_system-eq1}) and (\ref{new_system-eq2}). Then,
from Eq. (\ref{free-energy-density}) with $\mu_0 =0$,\cite{footnote3} one finds their energy densities:
\be
{\mathcal E}_{\mbox{\scriptsize SP}}= -\frac{2}{\pi l^{2}}\left(Z +\frac{\hbar^{2}}{8ml^{2}}(I_1+2I_2+I_3) \right)
\label{edsp}
\ee
and
\be
{\mathcal E}_{\mbox{\scriptsize LP}}=-\frac{2}{\pi l^{2}}\left(\tilde{\Delta}_0  -  \frac{e^{2}d}
{\kappa l^{2}}+\frac{\hbar^{2}}{8ml^{2}}(I_1+2I_2+I_3)
\right)\,,
\label{edlp}
\ee
where the integrals $I_i$ are defined in Eq. (\ref{integralsI-i}) and
the dielectric constant $\kappa$ is a free parameter in this model.
\begin{figure}[ht]
\includegraphics[width=7.0cm]{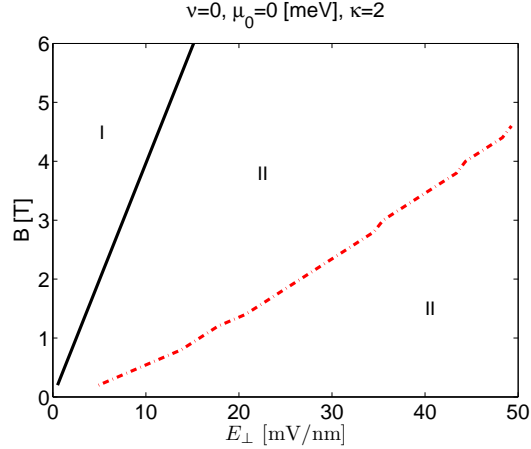}
\caption{The phase diagram for the $\nu=0$ QH state in the $(E_{\perp},B)$ plane.
Here the dielectric constant $\kappa=2$.}
\label{fig-nu0-phasediagram}
\end{figure}
Comparing the expressions for ${\mathcal E}_{\mbox{\scriptsize SP}}$ and
${\mathcal E}_{\mbox{\scriptsize LP}}$, one concludes
that there is a critical line separating the SP and LP
phases in the $\tilde{\Delta}_0 B$ plane:\cite{GGM1,GGM2}
\begin{equation}
\tilde{\Delta}_0^{\mbox{\scriptsize cr}} = Z + \frac{e^2 d}{\kappa l^2}.
\label{critical-line}
\end{equation}

In order to compare our results with experimental data in Refs.~\onlinecite{Weitz,Martin,SKim,Freitag},
it is convenient to express the top-bottom gates voltage imbalance $\tilde{\Delta}_0$ through the electric
field, $E_{\perp}= 2\tilde{\Delta}_0/ed$. Then, the critical line in the $E_{\perp}B$
plane takes the form
\begin{equation}
E_{\perp}^{\mbox{\scriptsize cr}}= \frac{2}{ed}\left(Z + \frac{e^2d}{\kappa l^2}\right).
\label{critical-line1}
\end{equation}
Using
$Z \simeq 0.67\,B[{\mbox T}]{\mbox K}$ and $l=\sqrt{\hbar c/eB}$, this relation
can be rewritten as
\begin{equation}
E_{\perp}^{\mbox{\scriptsize cr}}[\frac{{\mbox {mV}}}{{\mbox {nm}}}] \simeq (0.33 +
\frac{4.4}{\kappa})\, {B[{\mbox T}]}\,.
\label{critical-line-electric}
\end{equation}

We plot the phase diagram of the $\nu = 0$ state in Fig. \ref{fig-nu0-phasediagram}
(compare with Refs.~\onlinecite{GGM1,GGM2}). The I (II) area is that where the SP (LP)
solution is favorite. The red dashed line and the $B$ axis compose the boundary of the region where the
two solutions coexist: the solution I does not exist to the right of the (red) dashed line in the
region II. The black bold line is a critical line between the phases I and II. For balanced bilayer
($E_{\perp} = 0$), the SP solution is favorite due to the presence of the Zeeman term. At fixed $B$ for
sufficiently large electric field $E_{\perp}$, the LP solution is realized.

The gap $\Delta_{\nu=0}$ for the Hall state $\nu=0$ is equal to the difference
between the lowest empty (positive energy) level and the highest filled (negative energy)
level. For the SP solution, it is
\begin{equation}
\Delta_{\nu=0}=E_{1,1,+}-E_{-1, 1,-}=2\left[Z- \frac{ed}{2}\,E_{\perp} +\frac{\hbar^{2}}{2ml^{2}}
\left(I_{2}+I_{3}\right)\right],\quad E_{\perp} < \frac{2}{ed} [Z+\frac{\hbar^{2}}{2ml^{2}}
\left(I_{2}+I_{3}\right)],
\label{gap-nu-0-spin}
\end{equation}
and
\begin{equation}
\Delta_{\nu=0}=E_{-1,1,-}-E_{1, 1,+}=2\left[-Z+ \frac{ed}{2}\,E_{\perp}+\frac{\hbar^{2}}{2ml^{2}}
\left(I_{2}+I_{3}\right) -\frac{2e^{2}d}{\kappa l^{2}}\right],\quad E_{\perp} >
\frac{2}{ed} [Z-\frac{\hbar^{2}}{2ml^{2}}
\left(I_{2}+I_{3}\right)+\frac{e^{2}d}{\kappa l^{2}}]
\label{gap-nu-0-layer}
\end{equation}
for the LP solution.

At a fixed value of the magnetic field, the SP gap decreases with increasing $E_{\perp}$ up
to the point $E_{\perp} = 2[Z+\hbar^{2}/2ml^{2} \left(I_{2}+I_{3}\right)] /ed$, where it
vanishes.  On the other hand, the LP gap always increases with $E_{\perp}$. According to
the expression in Eq. (\ref{critical-line1}) for the critical line, the SP solution is favored for
$E_{\perp} < E_{\perp}^{\mbox{\scriptsize cr}} $, while at $E_{\perp} > E_{\perp}^{\mbox{\scriptsize cr}}$
the LP solution is realized. At the critical line, the gaps of the two phases coincide and are given by
\begin{equation}
\quad \Delta_{\nu=0}^{\mbox{\scriptsize cr}}=\frac{\hbar^{2}}{ml^{2}}\left(I_{2}+I_{3}-
\frac{2e^{2}dm}{\kappa\hbar^{2}}\right)\,.
\label{gap-across-critical-line}
\end{equation}

Let us describe the phase transition between the SP and LP phases in more  detail. In
Fig. \ref{fig-nu0-spectrumgap},
the gap $\Delta_{\nu=0}$ and the energy spectrum $E_{\xi n s}$ are shown as functions of
the electric field $E_{\perp}$ at the fixed value of the magnetic field $B = 2\,[{\mbox T}]$.
The critical value of $E_{\perp}$ is $E_{\perp}^{\mbox{\scriptsize cr}} \simeq 5.04\,{\mbox {mV/nm}}$ and
$\Delta_{\nu=0}^{\mbox{\scriptsize cr}} \simeq 3.73\,{\mbox {meV}}$ in this case.

\begin{figure}[ht]
\includegraphics[width=7.0cm]{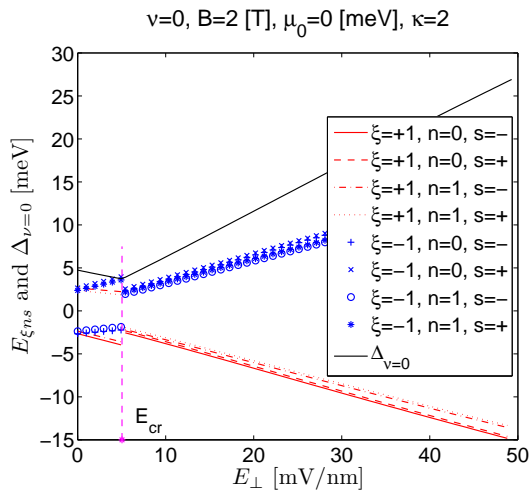}
\caption{The energy spectrum and the gap of the $\nu=0$ QH state
as functions of the electric field at the fixed magnetic field $B=2$ [T].
Here the dielectric constant $\kappa= 2.$}
\label{fig-nu0-spectrumgap}
\end{figure}

As one can see in Fig. \ref{fig-nu0-spectrumgap}, ignoring a small splitting of the LLs due
to the electric field, the symmetry of the SP phase at $E_{\perp} < E_{\perp}^{\mbox{\scriptsize cr}}$ is
$G_2 = U^{(K)}(1)_{+}\times U^{(K)}(1)_{-} \times U^{(K^{\prime})}(1)_{+} \times U^{(K^{\prime})}(1)_{-}
\times Z_{2V}^{(+)}\times Z_{2V}^{(-)}$ considered in Sec. \ref{2}.
Now, at the critical value $E_{\perp}^{\mbox{\scriptsize cr}}$, a jump in the energy spectrum takes place
(see Fig. \ref{fig-nu0-spectrumgap}).
Since the value of the gap $\Delta_{\nu=0}^{\mbox{\scriptsize cr}}$ at
$E_{\perp}=E_{\perp}^{\mbox{\scriptsize cr}}$ is nonzero, this leads to a
jump transformation of the $G_2$ symmetry into the $U^{(K)}(2)_S \times U^{(K^{\prime})}(2)_S$ one at
$E_{\perp} > E_{\perp}^{\mbox{\scriptsize cr}}$ (the LP phase), if the Zeeman term is ignored.
The presence of such a jump suggests that this phase transition is
a discontinuous (first order) one.\cite{order}

A noticeable feature of this transition is a kink singularity in the gap with a
minimum at $E_{\perp} = E_{\perp}^{\mbox{\scriptsize cr}}$ clearly seen in Fig. \ref{fig-nu0-spectrumgap}.
As will be discussed in Sec. \ref{5}, this fact is important
for understanding the behavior of the two-terminal conductance in the experiment in
Ref.~\onlinecite{Weitz}. It is interesting that as shown in Appendix, the last term
in the Hamiltonian (\ref{interaction1}), which is responsible for the bare splitting of the Landau
levels with $n=0$ and $n=1$, transforms the kink singularity into a (stronger) jump one. Although
for magnetic fields $B< 10$T, which are relevant for the present analysis, this jump is very small,
it could become relevant for dynamics with higher magnetic fields.

\subsection{The $\nu=1$ QH state}
\label{4b}

There are two main solutions at the $\nu=1$ filling factor (see Fig. \ref{fig-nu1-spectrumgap}).
The first solution is:
$$
\mbox{sgn} \left( E_{{1,0,-}} \right)=-1,\,\,\mbox{sgn} \left( E_{{-1,0,-}} \right)=-1,\,\,
\mbox{sgn} \left( E_{{1,1,-}} \right) =-1,\,\,\mbox{sgn} \left( E_{{-1,1,-}} \right)=-1,
$$
\begin{equation}
\mbox{sgn} \left( E_{{1,0,+}} \right)=-1,\,\,\mbox{sgn} \left( E_{{-1,0,+}} \right)=1,\,\,
\mbox{sgn} \left( E_{{1,1,+}} \right)=1,\,\,\mbox{sgn} \left( E_{{-1,1,+}} \right)=1\,.
\label{ferromagnetic-pf}
\end{equation}
This solution is closely connected with the SP one in Eq. (\ref{ferromagnetic}).
It differs from the latter only in that its LL with $\xi=1$, $n=0$, $s=+$ is filled.
It would be appropriately to call this solution a partially spin polarized (PSP) one.

The second solution is:
$$
\mbox{sgn} \left( E_{{1,0,-}} \right)=-1,\,\,\mbox{sgn} \left( E_{{1,0,+}} \right)=-1,
\mbox{sgn} \left( E_{{1,1,-}} \right)=-1,\,\,\mbox{sgn} \left( E_{{1,1,+}} \right)=-1,
$$
\begin{equation}
\mbox{sgn} \left( E_{{-1,0,-}} \right)=-1,\,\,\mbox{sgn} \left( E_{{-1,0,+}} \right)=1,\,\,
\mbox{sgn} \left( E_{{-1,1,-}} \right)=1,\,\,\mbox{sgn} \left( E_{{-1,1,+}} \right) =1.
\label{layer-asymmetric-pf}
\end{equation}

It is closely connected with the LP solution in Eq. (\ref{layer-asymmetric}). The difference is in that
the LL with $\xi=-1$, $n=0$, $s=-$ is now filled. It would be appropriate to call this solution
a partially layer polarized (PLP) one.

The energy densities of the partially filled spin polarized and layer polarized solutions are given by:
\be
{\mathcal E}_{\mbox{\scriptsize PSP}} =  -\frac{1}{2\pi l^{2}}\left[ 3 Z +\frac {\hbar^{2}}
{2ml^{2}}(I_1+ I_2+I_3)-\frac{e^{2}d}{4\kappa l^{2}}  +
\frac{eE_{\perp}d}{2} \right]
\ee
and
\be
{\mathcal E}_{\mbox{\scriptsize PLP}} =  -\frac{1}{2\pi l^{2}}\left[Z+\frac {\hbar^{2}}
{2ml^{2}}(I_1+ I_2+I_3)-\frac{9e^{2}d}{4\kappa l^{2}}
+\frac{3eE_{\perp}d}{2}\right]\,,
\ee
respectively. Comparing these two energy densities, we conclude that the critical line between the PSP
and PLP solutions exactly coincides with that between the SP and LP ones
(see Eq. (\ref{critical-line-electric})). It is quite noticeable that
this result is in complete accord with experimental data in Ref.~\onlinecite{Weitz} according
to which the phase transition between the two $\nu=+1$ states takes place near the same electric field
at which the  transition between the two $\nu=0$ states is observed.

\begin{figure}[ht]
\includegraphics[width=7.0cm]{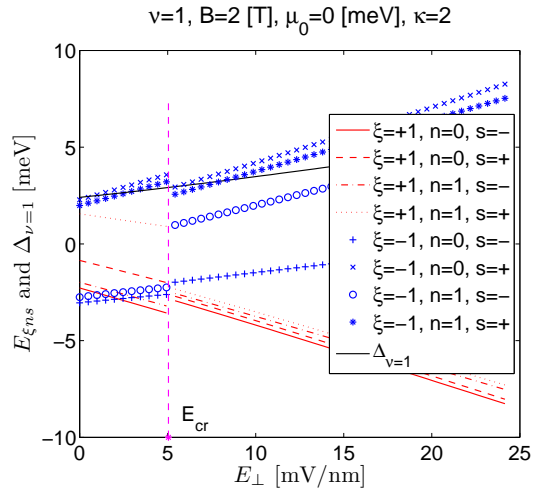}
\caption{ The energy spectrum and the gap of the $\nu=1$ QH state
as functions of the electric field at the fixed magnetic field $B=2$ [T].
Here the dielectric constant $\kappa= 2.$}
\label{fig-nu1-spectrumgap}
\end{figure}

The gap $\Delta_{\nu=1}$ and the energy spectrum $E_{\xi n s}$ for the $\nu=1$ QH state are
shown in Fig. \ref{fig-nu1-spectrumgap}. As one can see, a jump in the spectrum at
$E_{\perp} = E_{\perp}^{\mbox{\scriptsize cr}} \simeq 5.04 {\mbox {mV/nm}}$
is similar to that in the $\nu = 0$ QH state.
The phase transition at $E_{\perp} =E_{\perp}^{\mbox{\scriptsize cr}}$ corresponds to the transformation of the
$U^{(K)}(1)_{+}\times U^{(K)}(1)_{-} \times U^{(K^{\prime})}(1)_{+} \times
U^{(K^{\prime})}(1)_{-} \times Z^{(-)}_{2V}$ symmetry in the PSP phase into the
$U^{(K)}(2)_S \times U^{(K^{\prime})}(1)_{+}\times U^{(K^{\prime})}(1)_{-}$ in the PLP one
(as in the $\nu =0$ case, we ignore small splittings in the spectrum due to the Zeeman
term and the electric field).

As to the energy gap, there is an essential difference in its behavior in comparison
with the $\nu = 0$ gap.
The energy gap for the PSP solution is found to be
\be \Delta_{\nu=1}=E_{1,1,+}-E_{1,0,+}=\frac{\hbar^2}{2ml^2}(I_1+I_3-2I_2)
\label{gap1}\ee
in the region $E_{\xi n s} <  E_{\perp}^{\mbox{\scriptsize cr}}$, where it is the ground state.
In the region $E_{\xi n s} >  E_{\perp}^{\mbox{\scriptsize cr}}$, where the PLP solution is the
ground state, the energy gap takes the same form:
\be \Delta_{\nu=1}=E_{-1,1,-}-E_{-1,0,-}=\frac{\hbar^2}{2ml^2}(I_1+I_3-2I_2).
\label{gap11}
\ee
As a result, the gap in the ground state is a smooth function of $E_{\perp}$ (without a kink),
unlike the $\nu = 0$ case.
As will be discussed in Sec. \ref{5}, this feature and the phase transition at
$E_{\perp} =E_{\perp}^{\mbox{\scriptsize cr}}$
are important for understanding the behavior of the conductance at $\nu = 1$.

Note that both gap (\ref{gap1}) and gap (\ref{gap11}) are expressed as differences of the
LL energies with different orbital number $n$. This takes place also for the $\nu = 3$ QH state
(see Sec. \ref{4d} below). The polarization of the LLs states with different orbital numbers at odd-integer
filling factors was predicted in Ref.\onlinecite{Bar} based on the Hund's rule.

The following remark is in order. In addition to the two solutions described above,
there are also two other, marginal, solutions. The first of them is the same
as the PSP solution (\ref{ferromagnetic-pf}) except that instead the state with $\xi=1$, $n=0$, $s=+$,
it has the state with $\xi=1$, $n=1$, $s=+$ being filled.
The second marginal solution is the same as the PLP solution
(\ref{layer-asymmetric-pf}) except that the state with $\xi=-1$, $n=1$, $s=-$ is filled
instead that with $\xi=-1$, $n=0$, $s=-$. These marginal solutions have higher
energy density than solutions (\ref{ferromagnetic-pf}) and (\ref{layer-asymmetric-pf}),
however, they exist in a slightly larger region. Therefore, the marginal solutions can describe
the ground state of the system only in a small region. This issue will be considered in detail
elsewhere.

At last, we comment on how the bare splitting of the Landau levels with orbital
indices 0 and 1, discussed in Appendix, influences the $\nu=1$ solution.
First of all, the value of the critical electric field in the $\nu=1$ state remains the same as
in the $\nu=0$ one. Also, as in the $\nu=0$ case, the bare splitting leads
to a jump in the $\nu=1$ gap at the critical point. Its value coincides with that
in the $\nu=0$ gap. Although the jump is very small for magnetic fields $B< 10$T, it
could be relevant for dynamics with stronger magnetic fields.

\subsection{The $\nu=2$ QH state}
\label{4c}

There is only one solution that describes the ground state at the filling factor $\nu=2$
(see Fig. \ref{fig-nu2-spectrumgap}):
$$
\mbox{sgn} \left( E_{{1,0,-}} \right)=-1,\,\,\mbox{sgn} \left( E_{{-1,0,-}} \right) =-1,\,\,
\mbox{sgn} \left( E_{{1,1,-}} \right) =-1,\,\,\mbox{sgn} \left( E_{{-1,1,-}} \right) =-1,
$$
\begin{equation}
\mbox{sgn} \left( E_{{1,0,+}} \right) =-1,\,\,\mbox{sgn} \left( E_{{1,1,+}} \right) =-1,\,\,
\mbox{sgn} \left( E_{{-1,0,+}} \right) =1,\,\,\mbox{sgn} \left( E_{{-1,1,+}} \right) =1\,.
\label{nu-2}
\end{equation}

It can be obtained from the PSP solution (\ref{ferromagnetic-pf}) in the $\nu=1$ QH state
by filling the LLs with $\xi=1$, $n=1$, $s=+$ or, alternatively, from the PLP solution (\ref{layer-asymmetric-pf})
by filling the LLs with $\xi=-1$, $n=1$, $s=-$. In fact, perhaps more clear understanding of this solution can be
obtained by considering the SP and LP solutions in the $\nu=0$ QH state. For the SP solution (\ref{ferromagnetic}),
its LLs are spin polarized and the LLs with $s=-$ are filled. For the LP solution (\ref{layer-asymmetric}),
its LLs are layer polarized and the LLs with $\xi=1$ are filled. Solution (\ref{nu-2}) can be obtained from the SP
one by filling the LLs with $\xi=1$ or, alternatively, from the LP solution by filling the LL with $s=-$. It
would be appropriate to call it a partially spin-layer polarized (PSLP) solution.
The energy density of this solution is
\be
{\mathcal E}_{\nu=2} = -\frac{1}{\pi l^{2}}\left[Z +\frac {\hbar^{2}}{4ml^{2}}(I_1+2I_2+I_3)-
\frac{e^{2}d}{2\kappa l^{2}}+\frac{eE_{\perp}d}{2}\right].
\label{ed-2}
\ee
\begin{figure}[ht]
\includegraphics[width=7.0cm]{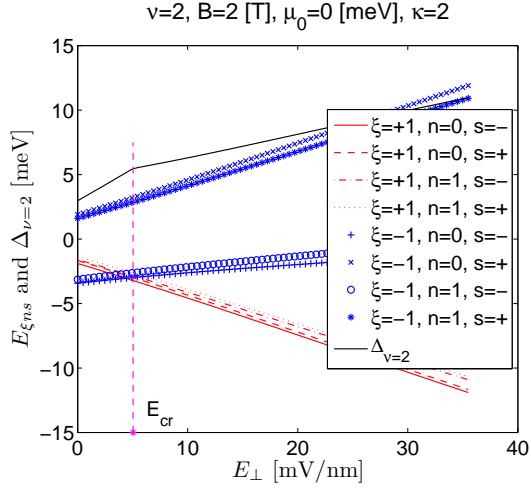}
\caption{ The energy spectrum and the gap of the $\nu=2$ QH state
as functions of the electric field at the fixed magnetic field $B=2$ [T].
Here the dielectric constant $\kappa= 2.$}
\label{fig-nu2-spectrumgap}
\end{figure}

The gap $\Delta_{\nu=2}$ and the energy spectrum $E_{\xi n s}$ for the $\nu=2$ QH state are
shown in Fig. \ref{fig-nu2-spectrumgap}. The gap is
\be
\Delta_{E_{\perp} < E_{\perp}^{\mbox{\scriptsize cr}}}=E_{-1,1,+}-E_{1,1,+}= e E_{\perp}d+\frac{\hbar^{2}}{m l^{2}}
\left(I_{2}+I_{3}-\frac{2e^{2}dm}{\kappa\hbar^{2}}\right)
\label{gap2a}
\ee
at $E_{\perp} < E_{\perp}^{\mbox{\scriptsize cr}} \simeq 5.04 {\mbox {mV/nm}}$, and
\be \Delta_{E_{\perp} > E_{\perp}^{\mbox{\scriptsize cr}}}=E_{-1,1,+}-E_{-1,1,-}=
2Z+\frac{\hbar^{2}}{m l^{2}}\left(I_{2}+I_{3}\right)
\label{gap2b}
\ee
at $E_{\perp} > E_{\perp}^{\mbox{\scriptsize cr}}$. Note, that at $E_{\perp}=0$ the gap
$\Delta_{\nu=2}$ (\ref{gap2a})
is smaller than the gap $\Delta_{\nu=0}$ (\ref{gap-nu-0-spin}) due to the negative Hartree contribution.

As one can see, unlike the $\nu=0$ and $\nu=1$ QH states,
there is no jump in the $\nu=2$ energy spectrum. Instead, a LLs crossing takes place when the
electric field $E_{\perp}$ is between 3.45 ${\mbox {mV/nm}}$  and 6.41 ${\mbox {mV/nm}}$. For
all values of $E_{\perp}$, the symmetry of the $\nu=2$ QH state is
$U^{(K)}(2)_S \times U^{(K^{\prime})}(1)_{+}\times U^{(K^{\prime})}(1)_{-}$,
if small splittings in the spectrum due to the Zeeman term are ignored.

As in the $\nu=0$ QH state, there is a kink in the $\Delta_{\nu=2}$
gap at $E_{\perp} = E_{\perp}^{\mbox{\scriptsize cr}} \simeq 5.04 {\mbox {mV/nm}}$,
however, its structure is quite different:
While there is a minimum in $\Delta_{\nu=0}$ at $E_{\perp} = E_{\perp}^{\mbox{\scriptsize cr}}$,
there is no either minimum or maximum in $\Delta_{\nu=2}$ at this same point (compare
Figs.  \ref{fig-nu0-spectrumgap} and \ref{fig-nu2-spectrumgap}). As will be discussed in Sec. \ref{5},
this fact is important for understanding the behavior of the conductance at $\nu = 2$.

The influence of the bare splitting term of the $n=0$ and $n=1$ LLs on the $\nu=2$ state
is described in Appendix. This term leads to the crossing value  $E_{\perp}^{\mbox{\scriptsize cross}}$
that is different from the critical value  $E_{\perp}^{\mbox{\scriptsize cr}}$ in the $\nu=0$ and $\nu=1$ states,
although their difference is small for $B < 10$T. The kink singularity in the $\nu=2$ gap remains unchanged.

\subsection{The $\nu=3$ QH state}
\label{4d}

There is one main solution and two marginal ones in this case. The latter will be considered
elsewhere. As to the main solution, it is given by (see Fig. \ref{fig-nu3-spectrumgap})
\be
\mbox{sgn} \left( E_{{1,0,-}} \right) =-1,\mbox{sgn} \left( E_{{
1,0,+}} \right) =-1,\mbox{sgn} \left( E_{{1,1,-}} \right) =-1,\mbox{sgn}
\left( E_{{1,1,+}} \right) =-1,
\ee
\begin{equation}
 \mbox{sgn} \left( E_{{-1,0,-}} \right) =-1,\mbox{sgn} \left( E_{{-1,0,+}} \right) =-1,\mbox{sgn}
\left( E_{{-1,1,-}} \right) =-1,\mbox{sgn} \left( E_{{-1,1,+}} \right) =1\,.
\label{nu-3}
\end{equation}
It can be obtained from the PSLP solution (\ref{nu-2}) in the $\nu=2$ QH state
by filling the LLs with orbital $n=0$ and $\xi=-1$, $s=+$. Its energy density is
\be
{\mathcal E}_{\nu=3} = -\frac{1}{2\pi l^{2}} \left[Z+\frac{eE_{\perp}d}{2}+
\frac{\hbar^{2}}{2ml^{2}}(I_1+ I_2+I_3)-\frac{e^{2}d}{4\kappa l^{2}}
\right]\,.
\label{ed-3}
\ee
\begin{figure}[ht]
\includegraphics[width=7.0cm]{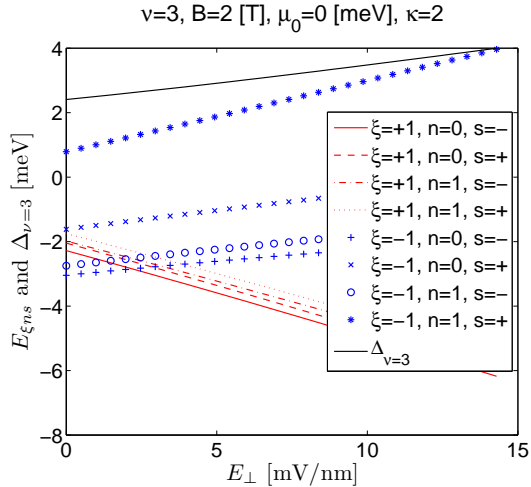}
\caption{ The energy spectrum and the gap of the $\nu=3$ QH state
as functions of the electric field at the fixed magnetic field $B=2$ [T].
Here the dielectric constant $\kappa= 2.$}
\label{fig-nu3-spectrumgap}
\end{figure}
The gap $\Delta_{\nu=3}$ and the energy spectrum $E_{\xi n s}$ for the $\nu=3$ QH state are
shown in Fig. \ref{fig-nu3-spectrumgap}. The gap is
\be \Delta_{\nu=3}=E_{-1,1,+}-E_{-1,0,+}= \frac{\hbar^{2}}{2m l^{2}}\left(I_{1}+I_{3}-2I_{2}\right)
\label{gap3}
\ee
for all values of $E_{\perp}$.

One can see in Fig. \ref{fig-nu3-spectrumgap} that as in the $\nu=2$ QH state,
there is no jump in the $\nu=3$ energy spectrum. Instead, a LLs crossing takes place when
the electric field $E_{\perp}$ is between 1.48 ${\mbox {mV/nm}}$ and 3.45 ${\mbox {mV/nm}}$.
The symmetry of the $\nu=3$ QH state is the same as that of the $\nu=2$ one,
$U^{(K)}(2)_S \times U^{(K^{\prime})}(1)_{+}\times U^{(K^{\prime})}(1)_{-}$
(if small splittings in the spectrum due to the Zeeman term are ignored).
The important difference between these two QH states is that the $\Delta_{\nu=3}$ gap in
Eq. (\ref{gap3}) is expressed as a difference of the LL energies with different orbital number $n$
(similarly to the gaps of the $\nu=1$ state).
Therefore we will call this solution a partially spin-layer-orbital (PSLOP) polarized one.

As in the case of the $\nu=1$ QH state, there is no kink in the gap $\Delta_{\nu=3}$.
This fact is relevant for understanding the behavior of the $\nu=3$ conductance discussed in Sec. \ref{5}.
As shown in Appendix, the influence of the bare splitting term of the $n=0$ and $n=1$ LLs
on the $\nu=3$ state is reduced to adding a small (for $B < 10$T) term in the gap.

\subsection{The $\nu=4$ QH state}
\label{4e}

The ground state is determined by the following solution in this case:
$$
\mbox{sgn} \left( E_{{1,0,-}} \right) =-1,\,\,\mbox{sgn} \left( E_{{1,0,+}} \right) =-1,\,\,
\mbox{sgn} \left( E_{{1,1,-}} \right) =-1,\,\,\mbox{sgn} \left( E_{{1,1,+}} \right) =-1,
$$
\begin{equation}
\mbox{sgn} \left( E_{{-1,0,-}} \right) =-1,\,\,\mbox{sgn} \left( E_{{-1,0,+}} \right) =-1,\,\,
\mbox{sgn} \left( E_{{-1,1,-}} \right) =-1,\,\,\mbox{sgn} \left( E_{{-1,1,+}} \right) =-1\,,
\label{nu-4}
\end{equation}
where the last empty LL in the PSLOP solution (\ref{nu-3}) (with $\xi=-1, n=1, s=+$)
is now being filled. Thus, the $\nu=4$ QH state describes the
filled LLL. The energy density of this solution equals
\be
{\mathcal E}_{\nu=4} =-\frac {\hbar^{2}}{4\pi ml^{4}}(I_1+2I_2+I_3 )\,.
\label{ed-4}
\ee
The energy spectrum $E_{\xi n s}$ for the $\nu=4$ QH state is
shown in Fig. \ref{fig-nu4-spectrumgap}. At $E_{\perp}=0$, its symmetry is
$G = U^{(K)}(2)_S \times U^{(K^{\prime})}(2)_S\times Z_{2V}^{(+)}\times Z_{2V}^{(-)}$
(if small splittings in the spectrum due to the Zeeman term are ignored). At nonzero
$E_{\perp}$, the symmetry becomes $U^{(K)}(2)_S \times U^{(K^{\prime})}(2)_S$.
Obviously, this QH state describes the gap between the LLL and the $n=2$ LL.
\begin{figure}[ht]
\includegraphics[width=7.0cm]{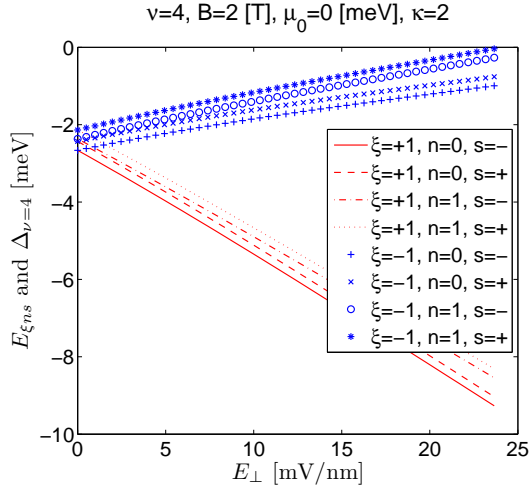}
\caption{ The energy spectrum of the $\nu=4$ QH state as a function of the electric field at
the fixed magnetic field $B=2$ [T].
Here the dielectric constant $\kappa= 2.$}
\label{fig-nu4-spectrumgap}
\end{figure}

\section{Comparison with experiment}
\label{5}


In this section, we will compare the results of our analysis with the recent experimental data in
Refs.~\onlinecite{Weitz,Martin,Freitag,SKim}. The comparison is hampered by the presence of disorder
in real bilayer samples, which is ignored in the present analysis. Still, as will be shown below,
the results of the analysis reproduce correctly the main characteristics of the experimental data.

Let us start from the critical line between the SP and LP solutions in
Eq. (\ref{critical-line-electric}). One can see that the maximum value
of its slope is obtained at the smallest permissible value $\kappa=1$ for the dielectric constant:
it is about $4.7 {\mbox {mV}}{\mbox {nm}^{-1}} {\mbox T^{-1}}$. On the other hand, the value of the slope
in experiments\cite{Weitz,SKim,Freitag} is about $11 {\mbox {mV}}{\mbox {nm}^{-1}} {\mbox T^{-1}}$.
This discrepancy may have its roots in disorder, depending on a type of the latter. For example,
an external electric field is more effective in a clean sample (considered in our model) and,
therefore, its critical value should be smaller than that in a real sample with charged
impurities. On the other hand, neutral impurities might act just in opposite direction by
diminishing the role of the Hartree interaction
in the gap equations, thus favoring the LP solution. This problem deserves further study.

Let us now turn to the energy gaps as functions of magnetic field at $E_{\perp} = 0$ shown in
Fig. \ref{fig-gaps}. Comparing this figure with Fig. 2 in Ref. \onlinecite{Martin}, one can see that
the hierarchy of the gaps is the same: $\Delta_{\nu=0} > \Delta_{\nu=2} >
\Delta_{\nu=1}=\Delta_{\nu=3}$. Moreover, the theoretical and experimental values of
$\Delta_{\nu=0}$ are close. The worst description in the model takes place for the smallest
$\Delta_{\nu=1}$ gap: while the experimental $\Delta_{\nu=1}$ is significantly
(by factor $10$) less than the experimental $\Delta_{\nu=0}$, the gap $\Delta_{\nu=1}$
is twice less than $\Delta_{\nu=0}$ in our model.
 This fact can be related to the observation in Ref. \onlinecite{Martin} that the QH states with
higher filling factors are subject to additional sources of disorder.
Then, it is not unreasonable to assume that the smallest gap, $\Delta_{\nu=1}$, is most sensitive to
disorder and therefore suppressed stronger. As to the $\nu=4$ QH state, it
describes a gap between the LLL and the $n=2$ LL, which is $\Delta_{\nu = 4} = \sqrt{2}\hbar\omega_c$,
up to small corrections.
\begin{figure}[ht]
\includegraphics[width=7.0cm]{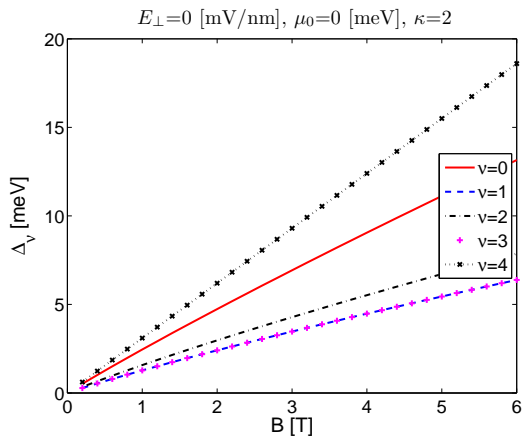}
\caption{The energy gaps as functions of magnetic field for different filling factors
at zero electric field and $\kappa=2$.}
\label{fig-gaps}
\end{figure}
In the rest of this section, we compare the behavior of the energy gaps as functions of electric field
in Figs. \ref{fig-nu0-spectrumgap} - \ref{fig-nu3-spectrumgap}
with that of the two-terminal conductance in experiment in Ref. \onlinecite{Weitz}.

\subsection{Energy gaps and conductance: even $\nu$}
\label{5b}

One of the main experimental results in Ref. \onlinecite{Weitz} is that the two-terminal conductance
is quantized except at particular values of the electric field $E_\perp$. Let us start from the
states with even filling factors.

For the $\nu =0$ state, the experimental data are the following. The conductance in the
$\nu =0$ state is quantized except two values of $E_\perp$, which increase with $B$.\cite{Weitz}
Let us argue that
these two values correspond to $E_\perp = \pm E_{\perp}^{\mbox{\scriptsize cr}}$, where
$E_{\perp}^{\mbox{\scriptsize cr}}$ is the critical value at which
the phase transition between the SP and LP phases takes place (see Fig. \ref{fig-nu0-spectrumgap};
recall that for convenience, we consider only nonnegative values of $E_\perp$ in the analysis).
As one can see in this figure, the gap $\Delta_{\nu=0}$ has a maximum at $E_{\perp}=0$ and a
minimum at $E_{\perp}=E_{\perp}^{\mbox{\scriptsize cr}}$. This implies that while the conductivity
is suppressed both around zero $E_{\perp}$ and for electric fields larger than
$E_{\perp}^{\mbox{\scriptsize cr}}$, it is enhanced (has a maximum) at
$E_{\perp}=E_{\perp}^{\mbox{\scriptsize cr}}$. This picture agrees with that in Fig. 2C
in Ref. \onlinecite{Weitz}.

As to $E_{\perp}^{\mbox{\scriptsize cr}}$, while its experimental value is
$E_{\perp}^{\mbox{\scriptsize cr}} \simeq 29 {\mbox {mV}}{\mbox {nm}^{-1}}$ at
$B= 2.65$T\,,\cite{Weitz} it is $E_{\perp}^{\mbox{\scriptsize cr}}\simeq 6.6{\mbox {mV}}{\mbox {nm}^{-1}}$
at $B= 2.65$T and $\kappa=2$ in our model (see Sec. \ref{4a}). This discrepancy can be
traced to the discrepancy between the experimental and theoretical values of the slope of
the critical line, which in turn could be
caused by disorder in real graphene samples (see the discussion above). As one can read from
Eq. (\ref{critical-line-electric}), the slope of the critical line in our model is
$2.5 {\mbox {mV}}{\mbox {nm}^{-1}} {\mbox T^{-1}}$  at $\kappa=2$ while its experimental value is about
$11 {\mbox {mV}}{\mbox {nm}^{-1}} {\mbox T^{-1}}$, i.e., by factor $4.4$ larger.
Taking this factor into account, we find that it will increase $E_{\perp}^{\mbox{\scriptsize cr}}$ from
$6.6{\mbox {mV}}{\mbox {nm}^{-1}}$ to
$29 {\mbox {mV}}{\mbox {nm}^{-1}}$ that agrees with the experimental value.

For the $\nu =2$ state, the picture is quite different. In this case, the conductance quantization
is broken only at $E_{\perp}=0$.\cite{Weitz} It agrees with the following facts
(see Fig. \ref{fig-nu2-spectrumgap}): a) unlike $\nu =0$, there is only one phase determining
the ground state for $\nu =2$ ; b) for $\nu =2$ the gap at $E_{\perp}=0$ is approximately $30-40\%$
less than that for $\nu =0$ and monotonically increases with $E_{\perp}$;  c) as already
mentioned above, the disorder suppresses the $\nu =2$ gap stronger than the $\nu =0$ one.
Let us also note that as was observed in Ref. \onlinecite{Weitz}, in stronger magnetic fields,
$B \gtrsim 7.8$T, the $\nu = 2$ conductance is quantized even around $E_{\perp}=0$, i.e.,
at such large $B$ (and therefore $\Delta_{\nu=2}$), the disorder is not able to break
the conductance quantization. As to the $\nu=4$ state, which describes a large gap between
the LLL and the $n=2$ LL, its conductance is obviously quantized for all $E_{\perp}$.

\subsection{Energy gaps and conductance: odd $\nu$}
\label{5c}

The quantization of the conductance of the $\nu =1$ state is broken at the same
finite values of $E_{\perp}$ as those in the $\nu =0$ one.\cite{Weitz} It is also
broken around $E_{\perp}=0$. As to the $\nu =3$ state, the situation is similar to
that in the $\nu =2$ state: the quantization of its conductance is broken only around
$E_{\perp}=0$. It is noticeable that in order to see these effects in the
odd $\nu$ states, one can use larger values of magnetic field ($B=3.43$T in
Ref. \onlinecite{Weitz}). This suggests that the odd $\nu$ states are more sensitive to
disorder than those with even $\nu$.

Let us discuss how these experimental data are reflected in the present analysis. The results for the
$\nu =3$ state, seems to be clear: as in the $\nu =2$ one, in this case there is one phase describing
the ground state and its gap has a minimum only at $E_{\perp}=0$. The case of the $\nu =1$ state is more
subtle. As for $\nu =0$, there are two phases describing the ground state (the PSP and PLP ones).
Moreover, the critical line dividing these phases exactly coincides with that between the SP
and LP ones at $\nu=0$. This is in accordance with the experimental data.
However, unlike the $\nu =0$ state, there is no minimum in the gap
at $E_{\perp}^{\mbox{\scriptsize cr}}$ (see Fig. \ref{fig-nu1-spectrumgap}).
Instead, the gap is a smooth monotonically increasing function of $E_{\perp}$ and
has a minimum only at $E_{\perp}=0$. This seems to suggest
that, in disagreement with experiment, the quantization of the conductance at $\nu =1$
should be broken only at $E_{\perp}=0$, {\it despite the presence of the phase transition at}
$E_{\perp}=E_{\perp}^{\mbox{\scriptsize cr}}\simeq 5.04 {\mbox {mV}}/{\mbox {nm}}$.

How can one solve this puzzle? A possible explanation might be as follows. Such a smooth
behavior of the $\nu=1$ gap occurs due to  delicate cancelations of different terms
in calculating the gap in the PSP and PLP phases. This cancelation could be just
an artifact of the approximation used in the present
analysis. On the other hand, it may also suggest that although beyond this approximation
a dynamics responsible for the conductance quantization breakdown at
$E_{\perp}=E_{\perp}^{\mbox{\scriptsize cr}}$ at $\nu=1$ does exist, it is weaker than that at $\nu=0$,
{\it even in perfectly clean samples}. This point deserves further study.

\section{Conclusion}
\label{6}

In this paper, the analysis of Refs. \onlinecite{GGM1,GGM2} was extended beyond the neutral point
with the filling factor $\nu=0$ to describe the doped QH states with filling factors
$\nu=\pm 1, \pm 2, \pm 3$, and $\nu=\pm 4$. It is noticeable that such a relatively simple
model as the present one reproduces the main characteristics of  experimentally
observed broken-symmetry LLL states in a magnetic field. One important ingredient in
the model is a very strong screening of the Coulomb interactions described by the polarization
function. Such a strong screening radically changes the form of the interaction and is
responsible for a linear scaling of dynamical gaps with a magnetic field in bilayer
graphene in contrast to monolayer graphene where a $\sqrt{B}$ scaling takes place.
The behavior of gaps as functions of magnetic field at zero electric field for different
filling factors, Fig.\ref{fig-gaps}, is found to be in a good agreement with experimental
data. \cite{Martin} The amplitudes of the gaps at the $\nu = \pm 1, 3$ and $\nu = \pm 2$
plateaus are significantly smaller than the amplitude of the $\nu = 0$ gap, due to the
separate filling of the $n=0$ and $n=1$ orbital Landau levels and the negative contribution
of the Hartree term, respectively.

At nonzero electric field, we found a critical line in the plane of electric field,
$E_{\perp}$, and magnetic field, B, that separates both the spin
and layer polarized phases at $\nu=0$ and the partially spin and layer polarized phases at
$\nu=\pm1$. On the other hand, there are unique phases for $\nu = \pm 2$ and $\nu = \pm 3$
QH states.
The phase transition point moves out to larger electric fields as the magnetic field is increased, which
implies that the ferromagnetic phase is stabilized by a magnetic field and is
destabilized by an electric field.

By studying the evolution of the gaps with an external electric field, we revealed
a strong correlation between this evolution and
the behavior of the conductance in the experiment:\cite{Weitz}
the values of the electric field at the minima of the gaps correspond to those
ones where the conductance is not quantized. The only exception is the $\nu=\pm1$
QH state, where the gap has a minimum only at $E_{\perp}=0$ and not at finite values
of the electric field (possible reasons for such a disagreement were considered in Sec. \ref{5c}).

Finally, we would like to note that as has been recently shown in Ref. \onlinecite{Cserti},
nontrivial order parameters can also have important effects on the minimal dc conductivity in
bilayer graphene.

{\it Note added}: {After this work was submitted for publication, a new experimental
paper, Ref. \onlinecite{Velasco}, has appeared, in which  the phase diagram of the $\nu=0$ QH state
in bilayer graphene is studied. In agreement with the theoretical predictions,\cite{GGM1,GGM2,Nandkishore2,Falko}
the critical line observed in that work separates  the spin-polarized  and
layer-polarized states and has the form  similar to the theoretical one, except the region with very
low values of $B$, where the LLL approximation is apparently unreliable.
In another recent experimental paper, \cite{Mayorov} a phase transition to the nematic state was observed
in bilayer graphene without magnetic field. The nematic state has been predicted in theoretical
works.\cite{nematic} It breaks the rotational symmetry and keeps quasiparticles gapless. It would
be interesting to study a competition between the gapped state and the nematic state due to
the Coulomb interaction and other interactions in bilayer graphene in a magnetic field.
}

\begin{acknowledgments}
V.A.M. acknowledges useful discussions with Giovanni Fanchini.
The work of E.V.G. and V.P.G. was supported partially by the Scientific Cooperation
Between Eastern Europe and Switzerland (SCOPES) programme under Grant
No.~IZ73Z0-128026 of the Swiss National Science Foundation (NSF),  the European
FP7 program, Grant No. SIMTECH 246937, and by the joint Ukrainian-Russian SFFR-RFBR Grant No.~F40.2/108.
V.P.G. acknowledges a collaborative grant from the
Swedish Institute. The work of J.J. and V.A.M. was supported by the Natural
Sciences and Engineering Research Council of Canada.
\end{acknowledgments}

\appendix

\section{Bare splitting of the Landau levels with $n=0$ and $n=1$}
\label{appendix}

In this Appendix, we comment on the splitting of the Landau levels with orbital
indices 0 and 1 due to the bias electric field which leads to the sombrero shape of the upper
band in the Hamiltonian of four-band model at zero magnetic field. \cite{McC,CN,Chakraborty}
Under the reduction to the two-band model, the influence of the high-energy band is taken into account
by the last term in the Hamiltonian (\ref{interaction1}). The bare splitting of the $n=0$
and $n=1$ LLs is given by the formula \cite{McC} $\delta=edE_{\perp}\hbar\omega_{c}/\gamma_{1}
\equiv \tilde{\Delta}_0\bar{\delta}$, where the dimensionless $\bar{\delta}$ is
$\bar{\delta}= 2\hbar\omega_c/\gamma_1 \approx 0.011 B$[T].

This term leads to a modification of the gap equations (\ref{new_system-eq1})
and (\ref{new_system-eq2}) for the LLL with $n=0$ and $n=1$:
\be \tilde{\Delta}_0\to(1-n\bar{\delta})\tilde{\Delta}_0\,.\ee
The free energy density (\ref{free-energy-density}) of the system will also receive a correction:
\be {\cal E}\to{\cal E}= -\frac{1}{8\pi l^2}
\sum_{\xi=\pm}\sum_{s=\pm}\sum_{n=0,1}\left[E_{\xi ns}
+\mu_0+sZ-\xi\tilde{\Delta}_0(1-n\bar{\delta})\right]\,\mbox{sgn}(E_{\xi ns})\label{A2}\,.\ee
Note that the values of the magnetic field are less than 10T in the experiment
with suspended bilayer graphene\cite{Weitz,Martin} the results of our analysis are compared with. Therefore
in this case the value of $\bar{\delta}=0.011B$[T] is less than 0.1, and all these corrections are small.
On the other hand, they could become relevant for stronger magnetic fields.

By combining analytical and numerical methods, the modified gap equations and free energy density were
analyzed. Now we list the main results that were obtained in this analysis.

(1) For the $\nu=0$ state in Sec. \ref{4a}, the transformation (\ref{A2}) corresponds to
$E_\perp$ in the free energy density of the LP phase effectively
transforming as $E_\perp\to (1-\bar{\delta}/2)E_\perp$. Because of that, the
critical value $E_{\perp}^{\mbox{\scriptsize cr}}$ separating the SP and LP phases changes:
\be
E_{\perp}^{\mbox{\scriptsize cr}} = \frac{2}{ed(1-\bar{\delta}/2)}
\left(Z + \frac{e^2d}{\kappa l^2}\right).
\ee
The gaps for both the SP and LP phases are modified through the transformation
$E_\perp\to (1-\bar{\delta})E_\perp$:
\be \Delta_{\nu=0}^{\mbox{\scriptsize SP}} = \frac{\hbar^2}{ml^2}(I_2+I_3)+
2[Z - (1 - \bar{\delta}) eE_\perp d/2], \quad\quad
\Delta_{\nu=0}^{\mbox{\scriptsize LP}}=\frac{\hbar^2}{ml^2}(I_2+I_3) -2[Z +
\frac{2e^2d}{\kappa l^2} -(1 - \bar{\delta}) eE_\perp d/2]. \ee
As a result, at the critical point, the two
gaps will not coincide anymore. Instead, there is a jump in the gap there:
\be \Delta_{\nu=0}^{\mbox{\scriptsize SP}}(\tilde\Delta_0=\tilde\Delta_0^{\mbox{\scriptsize cr}} )
-\Delta_{\nu=0}^{\mbox{\scriptsize LP}}(\tilde\Delta_0
=\tilde\Delta_0^{\mbox{\scriptsize cr}})=\frac{2\bar{\delta}}{1-\bar{\delta}/2}
\left(Z+\frac{e^2d}{\kappa l^2}\right).
\label{A4}\ee
(2) For the $\nu=1$ state in Sec. \ref{4b}, the transformation (\ref{A2}) corresponds to $E_\perp$
effectively transforming as $E_\perp\to (1-\bar{\delta}/2)E_\perp$ in the free energy density
in both the PSP and PLP phases.
As a result, the critical value $E_{\perp}^{\mbox{\scriptsize cr}}$ is modified in the same way as for the
$\nu=0$ state and still coincides with the critical value in the latter. The gaps for the
PSP and PLP solutions are modified as:
\be
\Delta_{\nu=1}^{\mbox{\scriptsize PSP}} = \frac{\hbar^2}{2ml^2}(I_1+I_3-2I_2)+
\bar{\delta} eE_\perp d/2, \quad\quad
\Delta_{\nu=1}^{\mbox{\scriptsize PLP}}=\frac{\hbar^2}{2ml^2}(I_1+I_3-2I_2)-\bar{\delta} eE_\perp d/2.
\ee
Therefore there is also a jump in the gap at the critical point:
\be
\Delta_{\nu=1}^{\mbox{\scriptsize PSP}}(\tilde\Delta_0=\tilde\Delta_0^{\mbox{\scriptsize cr}} )
-\Delta_{\nu=1}^{\mbox{\scriptsize PLP}}(\tilde\Delta_0
=\tilde\Delta_0^{{\mbox{\scriptsize cr}} })=\frac{2\bar{\delta}}{1-\bar{\delta}/2}
\left(Z+\frac{e^2d}{\kappa l^2}\right).
\ee
It is noticeable that this jump coincides with that in the $\nu=0$ state (see Eq. (\ref{A4})).\\
(3) For the $\nu=2$ state, $E_\perp$ in the free energy density also effectively transforms
as $E_\perp\to (1-\bar{\delta}/2)E_\perp$.
There is a crossing of the energy levels $E_{1,1,+}$ and $E_{-1,1,-}$ in this state.
As was shown in Sec. \ref{4c}, the value of $E_{\perp}^{\mbox{\scriptsize cross}}$ coincides with
that of $E_{\perp}^{\mbox{\scriptsize cr}}$ in the $\nu=0$ and $\nu=1$ states, if the
$\bar{\delta}$ correction is ignored.
However, when this correction is included, these two values become different.
Instead, at the crossing point,
\be
E_{\perp}^{\mbox{\scriptsize cross}}= \frac{2}{ed(1-\bar{\delta})}\left(Z + \frac{e^2d}{\kappa l^2}\right).
\ee
The form of the gap for $E_\perp\leq E_{\perp}^{\mbox{\scriptsize cross}}$ changes as
\be
\Delta_{E_{\perp} < E_{\perp}^{\mbox{\scriptsize cross}}} = (1-\bar{\delta})e E_{\perp}d+\frac{\hbar^{2}}{m l^{2}}
\left(I_{2}+I_{3}-\frac{2e^{2}dm}{\kappa\hbar^{2}}\right).
\ee
The form of the gap for $E_\perp> E_{\perp}^{\mbox{\scriptsize cross}}$ remains unchanged. \
As a result, the kink singularity in the gap at the crossing point also does not change.\\
(4) For the $\nu=3$ state in Sec. \ref{4d}, the $E_\perp$ in the free energy density effectively
transforms again as
$E_\perp\to (1-\bar{\delta}/2)E_\perp$. The gap in this phase changes to
\be
\Delta_{\nu=3}=\frac{\hbar^{2}}{2m l^{2}}\left(I_{1}+I_{3}-2I_{2}\right) -\bar{\delta} e E_\perp d/2.
\ee


\begin{thebibliography}{99}
\bibitem{footnote}The effects of the long-range Coulomb interaction
on the shift of the energy levels and optical transitions in biased bilayer graphene
in a magnetic field have been previously studied in Ref.\onlinecite{Abergel}. In that paper,
in order to find the energy levels shift, the diagonalization of a many-body Hamiltonian with the Coulomb
interaction was performed by using the noninteracting many body basis in the Hilbert space
constructed from the single-particle states.
\bibitem{footnote1} This feature is directly connected with that
the valley and layer indices are equivalent in the LLL.
\bibitem{footnote3}
Note that because the gap between the filled and unfilled levels in the states with
$\nu=0, \pm1, \pm2, \pm3$, and $\pm4$ contains zero value of the energy for all
relevant values of the magnetic and electric fields
(see Figs. \ref{fig-nu0-spectrumgap} - \ref{fig-nu4-spectrumgap}), one can choose the value
$\mu_0 = 0$ in Eqs. (\ref{new_system-eq1}) and (\ref{new_system-eq2}).
\bibitem{order} In a first order phase transition, two phases coexist and are
perturbatively stable, i.e., they are at least metastable. The latter is still an open
issue for the $\nu=0$ phase transition in bilayer graphene.
\end{thebibliography}
\end{document}